\documentclass[twocolumn,showpacs,showkeys]{revtex4}

\begin{document}

\title{On the Recent Results of E835 Experiment at FNAL}

\author{A.A.~Arkhipov}

\affiliation{Institute for High Energy Physics\\
142284 Protvino, Moscow Region, Russia}
\date{January 8, 2004}

\begin{abstract}
In this note we briefly concern several recent results of E835
Experiment at FNAL that are discussed in the framework of unified
picture on hadronic spectra elaborated in our earlier works. It has
been established that new E835 Collaboration results provided an
additional excellent confirmation of our theoretical conception.
\end{abstract}



\maketitle


No doubt the year 2003 will enter the history of particle physics as
a year of fundamental discoveries. A series of new mesons have been
discovered whose properties are in a strong disagreement with the
predictions of conventional QCD-inspired quark potential models. New
narrow meson $D_{sJ}^{*+}(2317)$, decaying into $D_s^+\pi^0$ has been
observed by BABAR Collaboration \cite{1} in the first. This
observation was soon confirmed by CLEO Collaboration \cite{2}, which
have also established the existence of a new narrow state with a mass
near 2.46 GeV in its decay to $D_s^{*+}\pi^0$. Belle Collaboration
\cite{3} reported the first observation of the $D_{sJ}(2317)$ and
$D_{sJ}(2457)$ in $B$ decays: $B \rightarrow \bar DD_{sJ}(2317)$ and
$B \rightarrow \bar DD_{sJ}(2457)$ with a subsequent $D_{sJ}(2317)$
decay to $D_s\pi^0$ and $D_{sJ}(2457)$ decay to $D_s^*\pi^0$ and
$D_s\gamma$ final states. Both CLEO and Belle observations of
$D_{sJ}(2457)$ have recently been confirmed by BABAR \cite{4}.
Moreover, Belle Collaboration has recently reported \cite{5,6} the
discovery of very narrow $X(3872)$-meson state
($\Gamma_{X(3872)}^{tot}<2.3 MeV$) in the $J/\psi\pi^+\pi^-$
invariant mass distribution in the $B$ decay $B^{\pm} \rightarrow
K^{\pm}J/\psi\pi^+\pi^-$. This observation of Belle Collaboration was
confirmed not long since by CDF at Fermilab \cite{7}. The mass
measurement presented by CDF 3871.4 $\pm$ 0.7 $\pm$ 0.4 MeV is in
agreement with the result of Belle. It should be noted, in
particular, that the mass 2317 MeV is approximately 41 MeV below the
$DK$ threshold but the mass 3872 MeV is very near the $D^0{\bar
D}^{*0}$ threshold, while the $D^+D^{*-}$ channel with approximately
8 MeV higher threshold mass is forbidden for $X(3872)$ decay by phase
space. What is remarkable here is that all new narrow states have
been observed at the masses which are surprisingly far from the
predictions of conventional quark potential models. It is still more
remarkable that all new observed states are very narrow, their total
widths being about a few MeV. The small widths were found to be in
contradiction with quark model expectations. Does it mean the end of
the constituent quark model? In any case, this means either
considerable modifications in the conventional quark models have to
be introduced or that completely new approaches should be applied in
hadron spectroscopy.

In our previous papers \cite{8,9,10,11,12,13,14,15} we have claimed
that existence of the extra dimensions in the spirit of Kaluza and
Klein together with some novel dynamical ideas may provide new
conceptual issues for the global solution of the spectral problem in
hadron physics. In fact, we have performed an analysis of
experimental data on the mass spectra of the two-nucleon system,
two-pion and three-pion systems, strange mesons, charmed and
charmed-strange mesons, and found out that simple formula provided by
Kaluza-Klein approach with the fundamental scale calculated before
\cite{8} has excellently described the experimentally observed hadron
spectra. The results of this analysis are partially summarized in
Report \cite{16} presented at Xth International Conference on Hadron
Spectroscopy HADRON'03 (August 31 -- September 6, 2003,
Aschaffenburg, Germany) where it is shown, in particular, that all
new observed states mentioned above are excellently incorporated in
our theoretically developed conception.

Really, a wealth of new and exciting experimental data have been
presented at the Conference HADRON'03. We have touched some of new
experimental data in the Report \cite{16}. In this note we will
concern the recent results of E835 Experiment at Fermilab presented
in Aschaffenburg in the talks of Claudia Patrignani \cite{17} and
Ismail Uman \cite{18}.

In Table 1 from Ref. \cite{9} the theoretically calculated
Kaluza-Klein tower of KK-excitations for the two-nucleon system and
the experimentally observed mass spectra of proton-proton and
proton-antiproton systems above the elastic threshold have been
shown. In Ref. \cite{9} there are the references  where the
experimental data have been extracted from. It is non-trivial that
Kaluza-Klein scenario predicts $M_n^{pp}=M_n^{p\bar p}$ i.e. a
special sort of (super)symmetry between fermionic (dibaryon) and
bosonic states, and Table 1 contains an experimental confirmation of
this fact as well. There are the blanks in Table 1 which have to be
filled in the future experimental studies. It was pleased for us to
hear that E835 have precisely measured directly the mass and width of
$\eta_c(1{}^1S_0)$ in $\bar pp$ annihilation: $M(\eta_c)=2984.1\pm
2.1\pm 1.0$ MeV and $\Gamma(\eta_c)=20.4^{+7.7}_{-6.7}\pm 2.0$ MeV
\cite{17}. This new E835 measurement shown in Table~1 by bold-face
number just filled the $M_{28}^{p\bar p }$-storey of the Kaluza-Klein
tower.

New observations of $\bar pp \rightarrow \chi_0 \rightarrow
\pi^0\pi^0, \eta\eta$ through interference with the continuum and
precise measurements of the mass and width ($M(\chi_0)=3415.5\pm
0.4\pm 0.07$ MeV and $\Gamma(\chi_0)=10.1\pm 1.0$ MeV \cite{17}) are
also nice news for us. As is seen from Table 2 $\chi_0$-state just
occupy the $M_{39}^{\eta\eta}$-storey of Kaluza-Klein tower for
$\eta\eta$ system.

In the same Table 2 new (preliminary though) results of E835
Collaboration \cite{18} for the masses of resonances decaying into
$\eta\eta$ have been shown by bold-face numbers. New results of E835
Collaboration \cite{18} for the masses of resonances decaying into
$\eta\pi$ have been presented in Table 3 by bold-face numbers too.
Asterisks in Tables 2-3 mark the states which have not been seen
before. It was a great pleasure to establish that new E835
Collaboration results provided an additional excellent confirmation
of our theoretical conception \cite{16}.

In summary we would like once again to emphasize that the deep idea
of existence the extra dimensions received an additional experimental
confirmation. Here we have briefly concerned several recent results
of E835 Experiment at FNAL that were discussed in the framework of
unified picture on hadronic spectra elaborated in our works. It has
been shown that quite an interesting and exciting results of E835
Collaboration have excellently been included into unified picture.

This is certainly a remarkable fact that a series of our publications
was followed by the fundamental discoveries in hadron spectroscopy
mentioned above, and here we would like to point out that strong time
correlation as well.

It is clear that further experimental studies with a higher mass
resolution are of great importance. In particular, this refers to the
problem of a large resonance overlap. It will also be important to
learn how one could experimentally extract the fine structures in a
broad peak.  We believe that the idea of an ultra-high resolution
hadron spectrometer \cite{19} is a vital experimental problem which
can be solved in hadron physics in the nearest future. Anyway, it is
too much desirable, and I do hope our most courageous wishes will
come true.

\newpage
\begin{table*}
\caption{\label{tab:table1}Kaluza-Klein tower of KK excitations for
$pp(p\bar p)$ system and experimental data.}
\begin{ruledtabular}
\begin{tabular}{ccllccll}
n & $M_n^{pp}$\,MeV & $M_{exp}^{pp}$\,MeV & $M_{exp}^{p\bar p}$\,MeV
& n & $M_n^{pp}$\,MeV & $M_{exp}^{pp}$\,MeV & $M_{exp}^{p\bar
p}$\,MeV \\ \hline 1 & 1878.38 & 1877.5 $\pm$ 0.5 & 1873 $\pm$ 2.5 &
15 & 2251.68 & 2240 $\pm$ 5 & 2250 $\pm$ 15
\\
2 & 1883.87 & 1886 $\pm$ 1 & 1870 $\pm$ 10 & 16 & 2298.57 & 2282
$\pm$ 4 & 2300 $\pm$ 20  \\  3 & 1892.98 & 1898 $\pm$ 1 & 1897 $\pm$
1 & 17 & 2347.45 & 2350 & 2340 $\pm$ 40
\\
4 & 1905.66 & 1904 $\pm$ 2 & 1910 $\pm$ 30  & 18 & 2398.21 & & 2380
$\pm$ 10
\\
5 & 1921.84 & 1916 $\pm$ 2 & $\sim $ 1920 &  19 & 2450.73 &   & 2450
$\pm$ 10   \\
  &         & 1926 $\pm$ 2 &   & 20 & 2504.90 &   & $\sim$
  2500    \\
  &         & 1937 $\pm$ 2 & 1939 $\pm$ 2 & 21 & 2560.61 &
  &  \\
6 & 1941.44 & 1942 $\pm$ 2 & 1940 $\pm$ 1 & 22 & 2617.76 & & $\sim$
2620
  \\
  &         & $\sim$1945  & 1942 $\pm$ 5 & 23 & 2676.27 &  & \\
7 & 1964.35 & 1965 $\pm$ 2 & 1968 & 24 & 2736.04 & 2735 & 2710 $\pm$
20  \\
  &         & 1969 $\pm$ 2 & 1960 $\pm$ 15 & 25 & 2796.99 &  &
  \\
8 & 1990.46 & 1980 $\pm$ 2 & $1990^{\,+15}_{\,-30}$ & 26 & 2859.05 &
& 2850 $\pm$ 5 \\
  &         & 1999 $\pm$ 2 &  &  27 & 2922.15 &  &
\\
9 & 2019.63 & 2017 $\pm$ 3 & 2020 $\pm$ 3 & 28 & 2986.22 & &$\mathbf{2984 \pm 2.1 \pm 1.0}$ \\
 10 & 2051.75 & 2046 $\pm$ 3 & 2040 $\pm$ 40  & 29 & 3051.20 &
&
   \\
   &         & $\sim$2050 & 2060 $\pm$ 20 &  30 & 3117.04 &  &  \\
11 & 2086.68 & 2087 $\pm$ 3 & 2080 $\pm$ 10 &  31 & 3183.67 &  &  \\

   &         &            & 2090 $\pm$ 20 & 32 & 3251.06 &  &  \\
   &         & $\sim$2122 & 2105 $\pm$ 15 & 33 & 3319.15 &  & \\
12 & 2124.27 & 2121 $\pm$ 3 & 2110 $\pm$ 10 & 34 & 3387.90 &  & 3370
$\pm$ 10 \\
  &         & 2129 $\pm$ 5 & 2140 $\pm$ 30 & 35 & 3457.28
&  & \\  13 & 2164.39 & $\sim$2150 & 2165 $\pm$ 45 & 36 & 3527.25 & &
\\
   &         & 2172 $\pm$ 5 & 2180 $\pm$ 10 & 37 & 3597.77 &
   & 3600
$\pm$ 20         \\  14 & 2206.91 & 2192 $\pm$ 3 & 2207 $\pm$ 13 & 38
& 3668.81 &  &  \\
\end{tabular}
\end{ruledtabular}
\end{table*}

\newpage
\begin{table*}
\caption{\label{tab:table2} Kaluza-Klein tower of KK excitations for
$\eta\eta$ system and experimental data.}
\begin{ruledtabular}
\begin{tabular}{cccccc}
 n & $ M_n^{2\eta}$\,MeV &
 $M_{exp}^{2\eta}$\,MeV &  n & $ M_n^{2\eta}$\,MeV &
 $M_{exp}^{2\eta}$\,MeV \\ \hline
1  & 1097.74 & & 33 & 2948.47 &  \\
2  & 1107.10 & & 34 & 3025.66 &  \\
3  & 1122.54 & & 35 & 3103.15 &  \\
4  & 1143.80 & & 36 & 3180.91 &  \\
5  & 1170.56 & & 37 & 3258.94 &  \\
6  & 1202.47 & & 38 & 3337.19 &  \\
7  & 1239.11 & & 39 & 3415.68 & $\mathbf{\chi_{0}(3415.5\pm 0.4\pm 0.07)}$ \\
8  & 1280.10 & $f_2(1275)$ & 40 & 3494.36 &  \\
9  & 1325.01 & $\mathbf{2^{++}(1330 \pm 2)}$ & 41 & 3573.25 &  \\
10 & 1373.47 & & 42 & 3652.31 &  \\
11 & 1425.12 & & 43 & 3731.54 &  \\
12 & 1479.62 & $\mathbf{2^{++}(1477 \pm 5)}$ & 44 & 3810.93 &  \\
13 & 1536.66 & $f'_2(1525)$ & 45 & 3890.47 &  \\
14 & 1595.99 & $\pi_1(1600)$ & 46 & 3970.15 &  \\
15 & 1657.34 & & 47 & 4049.96 &  \\
16 & 1720.51 & $\mathbf{0^{++}(1734 \pm 4)}$ & 48 & 4129.90 &  \\
17 & 1785.29 & & 49 & 4209.95 &  \\
18 & 1851.53 & & 50 & 4290.11 &  \\
19 & 1919.07 & & 51 & 4370.38 &  \\
20 & 1987.78 & $\mathbf{4^{++}(1986 \pm 5)}$ & 52 & 4450.75 &  \\
21 & 2057.54 & $\mathbf{2^{++}(\sim 2030)}$ & 53 & 4531.21 &  \\
22 & 2128.24 & $\mathbf{2^{++}(2138 \pm 4)}$ & 54 & 4611.76 &  \\
23 & 2199.80 & & 55 & 4692.39 &  \\
24 & 2272.14 & & 56 & 4773.10 &  \\
25 & 2345.18 & $\mathbf{4^{++}(2352 \pm 8)^{*}}$ & 57 & 4853.89 &  \\
26 & 2418.86 & & 58 & 4934.75 &  \\
27 & 2493.13 & $\mathbf{6^{++}(2484 \pm 14)}$ & 59 & 5015.68 &  \\
28 & 2567.93 & & 60 & 5096.68 &  \\
29 & 2643.21 & & 61 & 5177.73 &  \\
30 & 2718.94 & & 62 & 5258.85 &  \\
\end{tabular}
\end{ruledtabular}
\end{table*}

\newpage

\begin{table*}
\caption{\label{tab:table3}Kaluza-Klein tower of KK excitations for
$\eta\pi$ system and experimental data.}
\begin{ruledtabular}
\begin{tabular}{cccc}
 n & $M_n^{\eta\pi^0}$\,MeV & $M_n^{\eta\pi^\pm}$\,MeV & $M_{exp}^{\eta\pi}$\,MeV \\
 \hline
1  & 690.08  & 694.47  &  \\
2  & 711.99  & 715.92  &  \\
3  & 744.86  & 748.26  &  \\
4  & 785.79  & 788.72  &  \\
5  & 832.74  & 835.28  &  \\
6  & 884.37  & 886.58  &  \\
7  & 939.76  & 941.73  &  \\
8  & 998.30  & 1000.05 &  \\
9  & 1059.49 & 1061.07 &  \\
10 & 1122.97 & 1124.40 &  \\
11 & 1188.40 & 1189.72 &  \\
12 & 1255.56 & 1256.78 &  \\
13 & 1324.22 & 1325.36 & $\mathbf{2^{++}(1330 \pm 2)}$ \\
14 & 1394.21 & 1395.27 & $1^{-+}(1400 \pm 20)$ \\
15 & 1465.36 & 1466.35 & $0^{++}(1474 \pm 20)$ \\
16 & 1537.54 & 1538.47 &  \\
17 & 1610.63 & 1611.51 &  \\
18 & 1684.53 & 1685.36 & $\mathbf{0^{++}(\sim 1700)}$ \\
19 & 1759.15 & 1759.94 & $\mathbf{2^{++}(1740 \pm 7)}$ \\
20 & 1834.42 & 1835.17 &  \\
21 & 1910.27 & 1910.98 &  \\
22 & 1986.64 & 1987.32 & $\mathbf{4^{++}(1986 \pm 5)}$ \\
23 & 2063.47 & 2064.12 &  \\
24 & 2140.73 & 2141.36 &  \\
25 & 2218.37 & 2218.97 & $\mathbf{4^{++}(2226 \pm 6)^{*}}$ \\
26 & 2296.36 & 2296.94 &  \\
27 & 2374.66 & 2375.22 &  \\
28 & 2453.25 & 2453.79 &  \\
29 & 2532.11 & 2532.63 &  \\
30 & 2611.21 & 2611.71 &  \\
\end{tabular}
\end{ruledtabular}
\end{table*}

\end{document}